\documentclass[11pt]{article}
\usepackage[colorlinks]{hyperref}

\usepackage{amsmath,amsfonts,amsthm,amssymb}
\usepackage{epsfig}  
\usepackage{subcaption}
\usepackage{feynmf}
\def\ii{\'\i}

\def\cao{\c c\~ao}

\def\ii{\'\i}

\def\cao{\c c\~ao}

\def\ftoday{{\sl {Le \number\day \space\ifcase\month 
\or janvier\or f\'evrier\or mars\or avril\or mai
\or juin\or juillet\or ao\^ut\or septembre\or octobre
\or novembre \or d\'ecembre\fi\space \number\year}}}    
\def\ptoday{{\sl {\number\day \space de\space \ifcase\month 
\or janeiro\or fevereiro\or mar{\c c}o\or abril\or maio
\or junho\or julho\or agosto\or setembro\or outubro
\or novembro \or dezembro\fi\space de\space \number\year}}}    
\def\gtoday{{\sl {Den \number\day. \ifcase\month 
\or Januar\or Februar\or M\"arz\or April\or Mai
\or Juni\or Juli\or August\or September\or Oktober
\or November \or Dezember\fi\space \number\year}}}    
\def\today{{\sl {\ifcase\month
\or January\or February\or March\or April\or May
\or June\or July\or August\or September\or October
\or November \or December\fi \space\number\day,\space 
                                            \number\year}}}

\newcommand{\g}{\gamma}           
\renewcommand{\d}{\delta}         \newcommand{\D}{\Delta}
\newcommand{\e}{\varepsilon}

\newcommand{\la}{\lambda}        
\newcommand{\m}{\mu}
\newcommand{\n}{\nu}
\newcommand{\om}{\omega}         \newcommand{\OM}{\Omega}
\newcommand{\p}{\psi}              
\newcommand{\s}{\sigma}           \renewcommand{\S}{\Sigma}

\newcommand{\CC}{{\cal C}}

\newcommand{\EE}{{\cal E}}

\newcommand{\PP}{{\cal P}}

\newcommand{\es}{\\[3mm]}

\newcommand{\sla}{\raise.15ex\hbox{$/$}\kern -.57em} 
\newcommand{\Sla}{\raise.15ex\hbox{$/$}\kern -.70em}

\def\Lp{\displaystyle{\biggl(}}
\def\Rp{\displaystyle{\biggr)}}

\newcommand{\lp}{\left(}\newcommand{\rp}{\right)}

\newcommand{\complex}{{\kern .1em {\raise .47ex
\hbox {$\scriptscriptstyle |$}}
    \kern -.4em {\rm C}}}
\newcommand{\real}{{{\rm I} \kern -.19em {\rm R}}}
\newcommand{\rational}{{\kern .1em {\raise .47ex
\hbox{$\scripscriptstyle |$}}
    \kern -.35em {\rm Q}}}
\renewcommand{\natural}{{\vrule height 1.6ex width
.05em depth 0ex \kern -.35em {\rm N}}}

\newcommand{\pa}{\partial}

\newcommand{\dsum}[2]{\displaystyle{\sum_{#1}^{#2}}}   
\newcommand{\dint}{\displaystyle{\int}}

\newcommand{\twiddle}{\lower.9ex\rlap{$\kern -.1em\scriptstyle\sim$}}

\newcommand{\equ}[1]{(\ref{#1})}
\newcommand{\eq}{\begin{equation}}
\newcommand{\eqn}[1]{\label{#1}\end{equation}}
\newcommand{\eea}{\end{eqnarray}}
\newcommand{\eqa}{\begin{eqnarray}}
\newcommand{\eqan}[1]{\label{#1}\end{eqnarray}}
\newcommand{\ba}{\begin{array}}
\newcommand{\ea}{\end{array}}
\newcommand{\eqac}{\begin{equation}\begin{array}{rcl}}
\newcommand{\eqacn}[1]{\end{array}\label{#1}\end{equation}}


\newcommand{\bz}{\begin{enumerate}}
\newcommand{\ez}{\end{enumerate}}

\begin{document}

\title{Behaviour of Charged Spinning Massless Particles} 

\author{Ivan Morales\footnote{Present address: 
Brasilian Center for Research in Physics,
CBPF, {BR-22290-180} Rio de Janeiro, Brazil.}, 
Bruno Neves, 
Zui Oporto\footnote{Present address: Carrera de F\ii sica, 
Universidad Mayor de 
San Andr\'es, La Paz, Bolivia.} and 
Olivier Piguet
\\[4mm]
{\small Departamento de F\ii sica, Universidade Federal de 
Vi\c cosa (UFV)}\\
{\small  Vi\c cosa, MG, Brazil}
}

\date{December 2017}    

\maketitle

\vspace{-5mm}

\begin{center}
{\small\tt E-mails:
mblivan@gmail.com, bruno.lqg@gmail.com, \\
azurnasirpal@gmail.com, opiguet@pq.cnpq.br }
\end{center}

\vspace{3mm}

\begin{abstract}

We revisit the classical theory of a  relativistic massless 
charged point particle with spin and interacting with an external 
electromagnetic field.
In particular, we give a proper definition of its kinetic energy and its total energy, 
the latter being conserved when the  external field is stationary. 
 {We also write the conservation laws for the linear and angular momenta.}
Finally, we find that the particle's velocity may differ from $c$ as a 
result of the 
spin---electromagnetic field interaction, without jeopardizing Lorentz invariance.

\end{abstract}


\tableofcontents

\section{Introduction}	

Although charged, massless particles have never been observed in 
the world of real particles, electrons in two-dimensional 
materials such as 
graphene behave as massless quasi-particles, i.e., they~show 
an approximately linear 
relativistic dispersion relation, of the type 
$E\sim v_{\rm F} |{\bf p}|$. 
The velocity $v_{\rm F}\sim 10^6$ m/s is the Fermi 
velocity, depending on the microscopic properties of the 
material. It plays the role of 
the velocity of light  for such a ``mini-relativistic theory''.
Moreover, beyond its (chiral) helicity, 
the quasi-electron possesses a quantum number, the ``pseudo-spin'',
which makes its  wave function have four components and obey
a massless Dirac equation. An introduction to the physics of graphene may be found in the reviews~\cite{graphene1,graphene2}.

{The existence of this very special behaviour justifies a 
re-visitation of the {extensive} literature dedicated to the 
theory of a classical, punctual, relativistic,
spinning particle~\cite{Frenkel1}--\cite{deriglazov2}, which may be massive or not, and charged or not, beginning with the pioneering work
of Frenkel~\cite{Frenkel1,Frenkel2}. The~generally accepted formalism using
odd Grassmann numbers in order to describe the spin degrees of freedom
was introduced by Martin~\cite{Martin} and then developed by the authors of Refs.~\cite{Berezin1}--\cite{vanholten-el-dyn}. It was already recognized in Ref.~\cite{Berezin1,Berezin2} that this is the most natural formalism, the Grassmann algebra turning itself to a Clifford algebra describing a spin one-half quantum particle upon canonical quantization. 
A~formalism based on a supersymmetry localized on
the world line of the particle, allowing for treating the massive as well as the massless particle, was introduced 
in Refs.~\cite{casalbuoni1}--\cite{brink2}. 
The formalism based on a Grassmann algebra seems also to be the only known way to derive the 
dynamics of a classical spinning particle, massive or not, 
from an action principle, as already indicated via suggesting examples
in Ref.~\cite{Martin}.
Interaction with
an external electromagnetic field is usually considered. 
Refs.~\cite{galvao-teitelboim,geyer} also treat the case of 
an external gravitational field.}

{A complete treatment of a particle that is both charged and
massless is still missing, although~many results may be found
in the literature quoted above. These concern above all 
the existence of an action principle. Our concern here 
is about the properties
of the solutions of the dynamical equations and about the 
possible conservation laws, namely those of energy,
linear momentum and angular momentum. 
Our first new result is that, in contrast with the case of 
the spinless particle,
 the~behaviour of the particle with non-zero spin
is drastically different what is expected from a massless particle:
its velocity may be lower or higher than the velocity of light $c$
($c=1$ throughout the paper).
Moreover, this happens without conflict with Special Relativity
(characterized by its fundamental parameter $c$):  
Lorentz covariance of the equations is always preserved. 
The second new result is about the conservation laws 
associated with the symmetry properties of the external field. 
In particular, the~Noether procedure applied to the case 
of a stationary external field allows us to obtain a 
proper definition of energy for a massless interacting 
particle---a definition that had not seemed to exist in 
the literature, up to now.
}

In order to be able to give 
a realistic classical interpretation, in the final part of this paper, we 
introduce, following~\cite{balachandran}, spin variables 
$\S_{\m\n}$ that are quadratic in the odd Grassmann variables
$\p_\m$, treating 
the $\S$ as real numbers, instead of even Grassmann numbers 
 that would be nilpotent. The resulting equations of motion 
are no more derivable from an action.
Moreover, they suffer incompatibilities, excepted for special 
external field configurations, e.g.,  for constant fields. 
It is for the latter configuration that we solve the equations, analytically or numerically, in order to gain some insight on the behaviours we mentioned above. 
{We must recall that various sets of dynamical equations are presented in the literature, based on an action principle or not.
We shall stick here to the set of equations derived in 
the Lagrangian formalism based on a Grassmann algebra 
as developed in~\cite{casalbuoni1}--\cite{brink2}}.

We restrict the scope of the present paper to the  interaction with
an {\it external} electromagnetic field. 
The specific problem of the radiation field has been treated 
by the authors of 
Refs.~\cite{kosyakov}--\cite{lechner}.

The plan of the paper is the following.
We begin in Section \ref{spinless} with a complete study of 
the spinless case in order to make 
some basic points more transparent. 
In Section \ref{spinning},
we present the results for the spinning case, with a last subsection
containing particular solutions of interest, 
 and finish with our conclusions, {together with a comparison with the known results for the massive particle}.

\section{The Spinless Relativistic Particle} \label{spinless}

The action for a classical spinless particle of mass $m$ of
electric charge $q$ interacting with an 
electromagnetic field, given by the potential vector 
$A_\m$, in four-dimensional Minkowski space-time 
may be written as the following integral on a time-like curve 
$\CC$ parametrized by $\la$~\cite{brink1,brink2,Kosyakov-book}:
\begin{equation}\label{action-spinless}
{S[x,e] }
= -\dint_{\!\!\!\CC} d\la\lp \dfrac{1}{2e(\la)}\dot{x}(\la)^2
+ \dfrac{e(\la)}{2} m^2 + q\, \dot{x}^\m(\la) A_\m (x(\la))\rp ,
\end{equation}
where\footnote{The units are defined by 
$c=\hbar=1$.
The Minkowski metric is 
$(\eta_{\m\n})$ = diag$(1,-1,-1,-1)$. 
The~dot means derivative with respect to $\la$ and
$\dot{x}^2$  stands for $\dot{x}^\m \dot{x}_\m$. Coordinates will also be denoted as $x^0=t$ and $(x^i,\,i=1,2,3)$ = $(x,y,z)$.}  
$x^\m(\la)$ are the coordinates of the particle's position and $e(\la)$
a real function on the curve $\CC$ parametrized by 
$\la$.
Under (infinitesimal) reparametrizations 
$\la'$ = $\la+\e(\la)$, the coordinates $x^\m$ transform as scalars 
and $e$ as a scalar density of weight 1:
\begin{equation}\label{repara}
\d x^\m = \e \dot{x}^\m,\quad \d e = \e\dot{e}+\dot{\e}e.
\end{equation}

Under these transformations, the action is invariant up to boundary terms, and the equations of motion  following from the variation of
$x^\m(\la)$,
\begin{equation}\label{eq-motion}
\dfrac{d}{d\la}\lp \dfrac{\dot{x}_\m}{e}\rp - q F_{\m\n}\dot{x}^\n
=0,
\end{equation}
where
\begin{equation}\label{F=dA}
F_{\m\n}=\pa_\m A_\n - \pa_\n A_\m,
\end{equation}
and the constraint following from the variation of $e(\la)$,
\begin{equation}\label{constraint}
\dfrac{\dot{x}^\m\dot{x}_\m}{e^2} = m^2  ,
\end{equation}
are covariant.

The propagation of a massless charged particle is described 
by the same action where $m$ is set to~zero. 

We observe that the constraint  \equ{constraint} is not 
completely independent of the equations of motion \equ{eq-motion}.
Indeed, multiplying the latter by $\dot{x}^\m/e$, we find that the left-hand side of {this constraint} is a constant:
\begin{equation}\label{constraint-stability}
\dfrac{d}{d\la}\lp \dfrac{\dot{x}^\m\dot{x}_\m}{e^2} \rp = 0 .
\end{equation}

This means that it will be sufficient to impose it at some initial value
of the parameter $\la$.

The solution of the constraint
 \equ{constraint} differs qualitatively 
in the massive and in the massless case. These cases will be
therefore treated separately in the following subsections.

The theory defined by the action \equ{action-spinless} can 
be considered as a gauge theory in the one-dimensional 
space-time defined by the world line $\CC$, the gauge 
invariance being that 
under reparametrizations~\equ{repara} and the fields being 
the position coordinates $x^\m(\la)$ and the ``\textit{einbein}''
 function $e(\la)$, the formers transforming as scalars and the 
latter as a density of weight 1. 

One way to fix the gauge is to fix a value for the non-physical
 variable $e(\la)$. One equivalent way is to simply choose 
 a particular parametrization, e.g.,  proper time or coordinate 
 time. Then $e(\la)$ will be determined by either the 
 constraint \equ{constraint}---if 
 $\dot{x}^\m\dot{x}_\m$ $\not=$ 0, i.e., in the massive case---
or the equations of motion \equ{eq-motion}.

\subsection{The Massive Case}

Let us begin with the proper time parametrization $\la$ = $\tau$. 
The 4-velocity $\dot{x}^\m$ then satisfies 
\[
\dot{x}^2=\dot{x}^\m\dot{x}_\m=1 ,
\]
so that the constraint \equ{constraint} solves for $e(\tau)$ as
\[
e=1/m ,
\]
where we have chosen the positive solution. The equation of motion
\equ{eq-motion} then takes the familiar covariant form
\[
m\ddot{x}_\m -  q\,F_{\m\n} \dot{x}^\n = 0,
\]
where the second term is the relativistic expression for the Lorentz
force.

In the time coordinate parametrization (the dot meaning now 
a time derivative), the 4-velocity takes the {form}
$(\g, \g \dot{\bf x})$, with $\dot{\bf x}$ = 
($\dot{x}^i$, $i=1,2,3$)
and $\g$ = $1/\sqrt{1- \dot{\bf x}^2}$. 
The constraint \equ{constraint}
solves now as 
\begin{equation}\label{constraint-massive}
\dfrac{1}{e} = m\g ,
\end{equation}
and the equations of motion \equ{eq-motion} read
\begin{equation}\label{eq-motion-massive}
\begin{array}{l}
m\dfrac{d}{dt}\g = q \bf{E} \cdot\dot{\bf x}, \vspace{3pt} \\
m\dfrac{d}{dt}\lp\g\dot{\bf x}\rp = q(\bf{E} 
+  \dot{\bf x} \times \bf{B})  ,
\end{array}
\end{equation}
where we have identified the electric and magnetic fields 
as 
\begin{equation}\label{el&mag-fields}
{\bf E}=(F_{01},F_{02},F_{03}) ,\quad 
{\bf B}=(-F_{23},-F_{31},-F_{12}) .
\end{equation}

In the stationary case, defined by $\pa_t A_\m=0$, we have 
a conserved energy obtained by integrating the first of 
{Equation} \equ{eq-motion-massive}:
\begin{equation}\label{en-cons-massive}
\EE = \dfrac{1}{e(t)}+q A_0(x(t)) = m\g(t)+q A_0(x(t)) ,
\end{equation}
where the integration constant $\EE$ is
the total energy.

\subsubsection{Example} 

{In the case of four-dimensional} space-time of coordinates 
$t,\,x,\,y,\,z$, with {\it constant fields} $\bf E$ = $(0,E,0)$ and
 $\bf B$ = $(0,0,B)$, 
we can perform a first integration of
{Equation} \equ{eq-motion-massive}, obtaining 
\begin{equation}\label{eq-motion-constB,E-massive}
\begin{array}{l}
m\g\dot{x} - q By + C_1 = 0 ,  \vspace{3pt} \\
m\g\dot{y} + qBx - qE t +C_2 = 0 ,  \vspace{3pt} \\
m\g\dot{z}+ C_3 = 0 ,
\end{array}
\end{equation}
$C_1$, $C_2$ and $C_3$ 
being integration constants.

\subsection{The Massless Case}\label{massless case}
\vspace{-6pt}
\subsubsection{Equations of Motion}
We are now going to investigate the main topics of this paper,
i.e., the motion of a massless charged particle in an 
electromagnetic field. The action is 
given in {Equation} \equ{action-spinless}, with now $m=0$.
The~main difference with respect to the massive one is 
in the constraint obtained by varying the variable $e(\la)$ 
in the action:
it now takes the form of the light-cone condition
\begin{equation}\label{l-c-constraint-cov}
\dot{x}^\m \dot{x}_\m = 0 ,
\end{equation}
and we see that, on the contrary of the massive case, 
it does not determine $e(\la)$.

The equations of motion are given by \equ{eq-motion} 
for a general parametrization. There is of course no 
proper time parametrization; we shall use the 
coordinate time as a parameter, so that they take the~form
\begin{equation}\label{eq-motion-massless}
\begin{array}{l}
\dfrac{d}{dt}\lp\dfrac{1}{e}\rp = q \bf{E} \cdot\dot{\bf x}  , \vspace{3pt} \\
\dfrac{d}{dt}\lp\dfrac{\dot{\bf x}}{e}\rp = q(\bf{E} 
+  \dot{\bf x} \times \bf{B})  .
\end{array}
\end{equation}

The constraint \equ{l-c-constraint-cov} now reads
\begin{equation}\label{l-c-constraint}
\dsum{i=1}{3}(\dot{x}^i)^2  = 1 .
\end{equation}

This constraint \equ{l-c-constraint} \textit{not} being taken 
into account, we have four differential equations for four functions,
 $x^i(t)$ and $e(t)$, 
of second order in the $x$s and first order in $e$. 
Its solutions depend therefore on seven integration constants. 
six of them can be fixed by six boundary conditions,
which may be chosen as six initial conditions at $t=0$:
\begin{equation}\label{initial-cond-massless}
x^i(0)=0,\ i=1,2,3, 
\quad (\dot{x}^1(0),\dot{x}^2(0),\dot{x}^3(0))=(v_{0x},v_{0y},v_{0z}) .
\end{equation}

Due to {Equation} \equ{constraint-stability}, the constraint 
\equ{l-c-constraint-cov} will be satisfied if it is verified at 
$t=0$, i.e.,
\begin{equation}\label{t=0-constraint}
(v_{0x})^2+(v_{0y})^2+(v_{0z})^2 = 1 .
\end{equation}

One of the integration constants remain free and will be discussed in Section \ref{energy}.

\subsubsection{Energy Equation}\label{energy}

The main difference with respect to the massive case is that  
the \textit{einbein} function $e(t)$ is no more determined 
by the constraint \equ{constraint}. Let us try to interpret it.
Its evolution is determined,
up~to an integration constant, by the first of {Equation}
\equ{eq-motion-massless}.
Restricting ourselves to the stationary case, where~$\pa_t A_\m=0$; hence, $\bf E$ = $\bf -\nabla A_0$, 
we see that this equation is a total time derivative, which yields
\begin{equation}\label{en-cons-massless}
\EE=\dfrac{1}{e(t)}+q A_0(x(t)) .
\end{equation}

The second term being the potential energy, we 
interpret the integration constant $\EE$ as the total 
energy of the particle, its ``kinetic energy'' being identified with
$1/e(t)$. In order to understand better the physical meaning of this,
let us normalize the electric potential---which in the 
stationary case is defined up to a constant---by 
\[
A_0(x(t)) = -\dint_{\!\!\!\!0}^t dt'\,{\bf E}(x(t'){\dot{\bf x}}(t') .
\]

In this situation, $\EE$ = $1/e(0)$, which may be interpreted 
as the kinetic energy accumulated until the time $t=0$. We shall assume $\EE$ to be positive:
\begin{equation}\label{energy-positivity}
\EE>0 .
\end{equation}

We may rewrite the second of {Equation} \equ{eq-motion-massless} as
\begin{equation}\label{eq-motion-massless'}
\dfrac{d}{dt}\lp(\EE-q A_0)\dot{\bf x}\rp = q(\bf{E} 
+  \dot{\bf x} \times \bf{B})
\end{equation}
and remark that the energy $\EE$---an arbitrary parameter---contributes to the inertia of the particle: increasing the value of $\EE$ implies more inertia.

\subsubsection{Example of a Constant Electromagnetic 
Field}\label{example_spinless}

In order to get more insight for the motion of the massless 
particle, let us consider {\it constant fields} \mbox{$\bf E$ = $(0,E,0)$} and  $\bf B$ = $(0,0,B)$, orthogonal to each other. 
We can perform a first integration of
{Equation}~\equ{eq-motion-massless'}, obtaining 
\begin{equation}\label{eq-motion-constB,E-massive'}
\begin{array}{l}
(\EE+qE y)\dot{x} - q By + C_1 = 0 , \vspace{3pt}\\
(\EE+qE y)\dot{y} + qBx - qE t +C_2 = 0 ,\vspace{3pt}\\
(\EE+qE y)\dot{z}+C_3 = 0 ,
\end{array}
\end{equation}
where the energy \equ{en-cons-massless} reads
\begin{equation}\label{energy'}
\EE = \dfrac{1}{e(t)}- q Ey(t).
\end{equation}

The integration constants $C_1$, $C_2$ and $C_3$ 
are fixed as
\begin{equation}\label{int-const'}
C_1=-\EE       v_{0x}   ,\quad C_2=-\EE v_{0y} ,\quad C_3=-\EE v_{0z} ,
\end{equation}
with $\sum_i (v_0^i)^2=1$, 
by the initial conditions \equ{initial-cond-massless}.

A peculiar feature of the solutions of {Equation}
\equ{eq-motion-constB,E-massive'} is a transition in their 
qualitative behaviour: for $|B|>|E|$, the trajectory is 
bounded in the $y$-direction, i.e., the direction of the 
electric field, whereas it is unbounded in the case 
$|B|<|E|$. In order to show this, let us solve the 
system \equ{eq-motion-constB,E-massive'} for $x$ and $y$:
\begin{equation}\label{trans_E=B}
\begin{array}{l}
x(t)=\dfrac{E}{B}t  + \dfrac{\EE v_{0y}}{qB} - \dfrac{\EE\dot{y}(B-Ev_{0x})}{qB(B-E\dot{x})} ,\es
y(t)=\dfrac{\EE(\dot{x}-v_{0x})}{q(B-E\dot{x})} .
\end{array}
\end{equation}

Since the velocity components are all bounded by 1 in 
absolute value, it is clear that, if
\mbox{$\vert B\vert$ $>$ $\vert E\vert$}, the denominator of 
the expression for $y(t)$ never vanishes, then $y(t)$ 
remains bounded. 
However, $x(t)$ is asymptotically linear in $t$ and thus 
is unbounded (unless the electric field vanishes). 

Solutions with $y(t)$ unbounded are those for  \mbox{$\vert B\vert$ $\le$ $\vert E\vert$}. 
This set includes the limiting  case \mbox{$\vert B\vert$ $=$ $\vert E\vert$}, 
where one explicitly checks that 
$y(t)$ and $x(t)$ go asymptotically as $x\sim t$ and  
$y\sim t^{2/3}$, respectively,
as $t$ $\to$ $\infty$, unless the initial velocity is 
transverse to the electric field: $(v_{0x},\,v_{0y},v_{0z})$ 
= $(1,\,0,\,0)$, in which case the solution of the equations---with the  given initial conditions \equ{initial-cond-massless}---is $x(t)=t$, $y(t)=0$.

\begin{figure}[htb]
\centering
\includegraphics[scale=0.5]{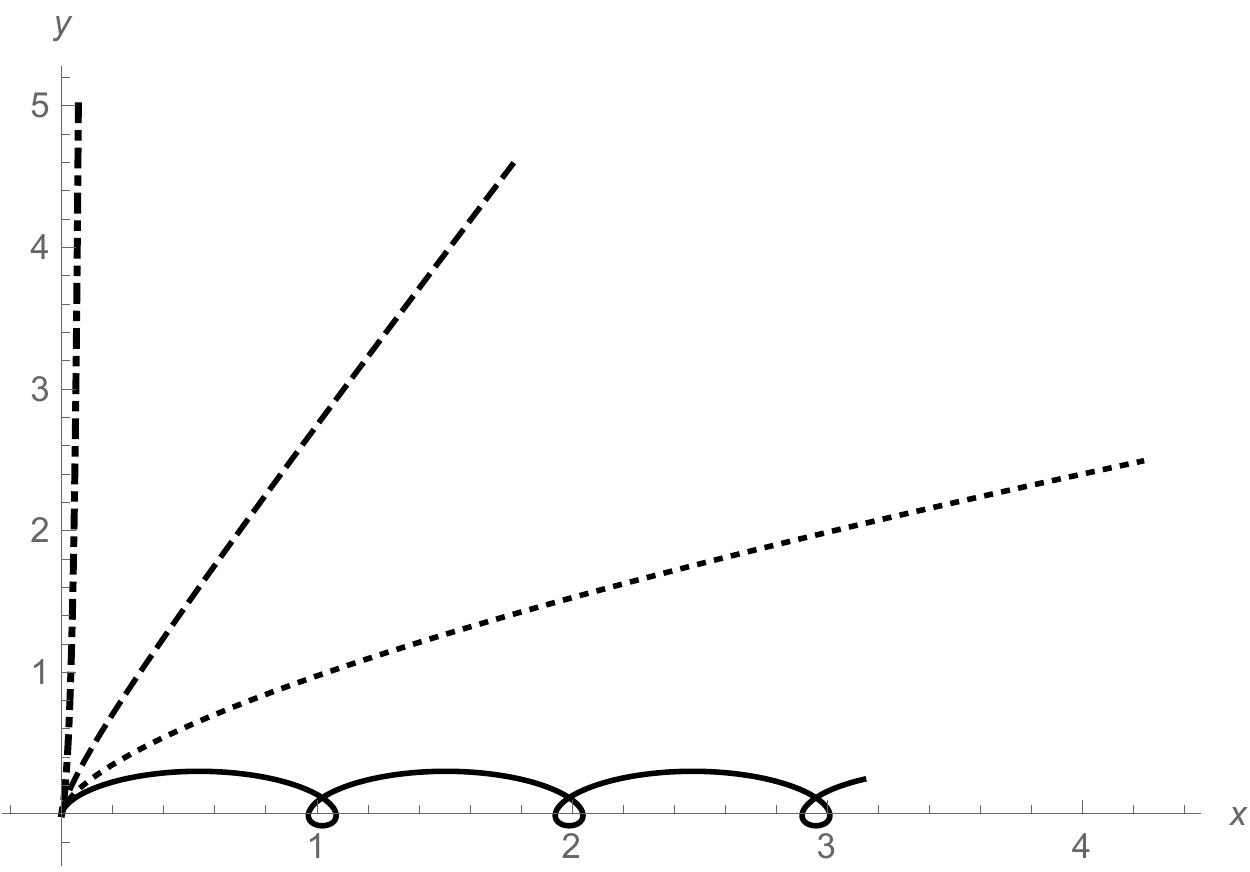}
\caption{\small 
Particle trajectories in the $z=0$ plane for 
$0\le t\le5$. Charge $q=1$, constant electric field $E$ in 
the positive $y$
direction, constant magnetic field $B$ in the positive $z$ direction.
Energy $\EE=0.2$, initial velocity ${\bf v}_0$ = $(0.1,0.995,0)$.
Solid line: $B=1.6,\, E=1$; dotted line: $B=E=1$;
dashed line: $B=0.4,\, E=1$; dotted-dashed line: $B=0,\, E=1$.
The $B=0$ trajectory would be on the upper vertical
axis in the case of ${\bf v}_0$ = $(0,1,0)$.}
\label{fig1}
\end{figure}

\begin{figure}[htb]
\centering
\begin{tabular}{cc}
\includegraphics[scale=0.55]{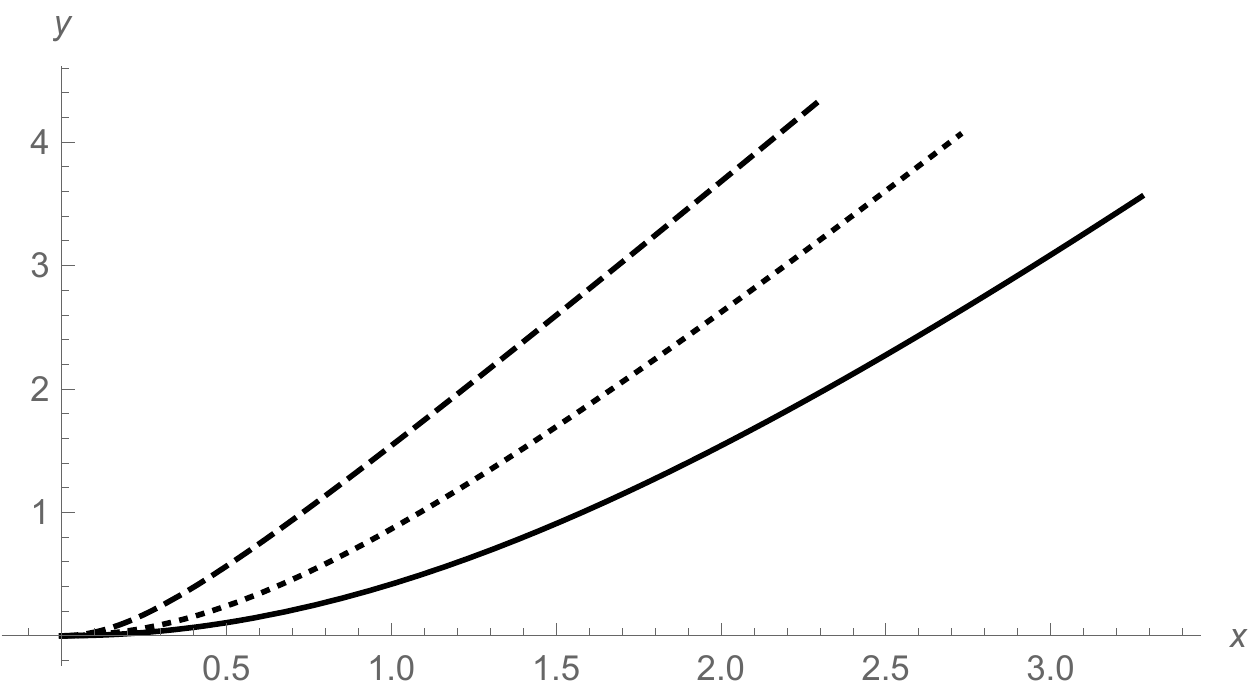}
&\includegraphics[scale=0.55]{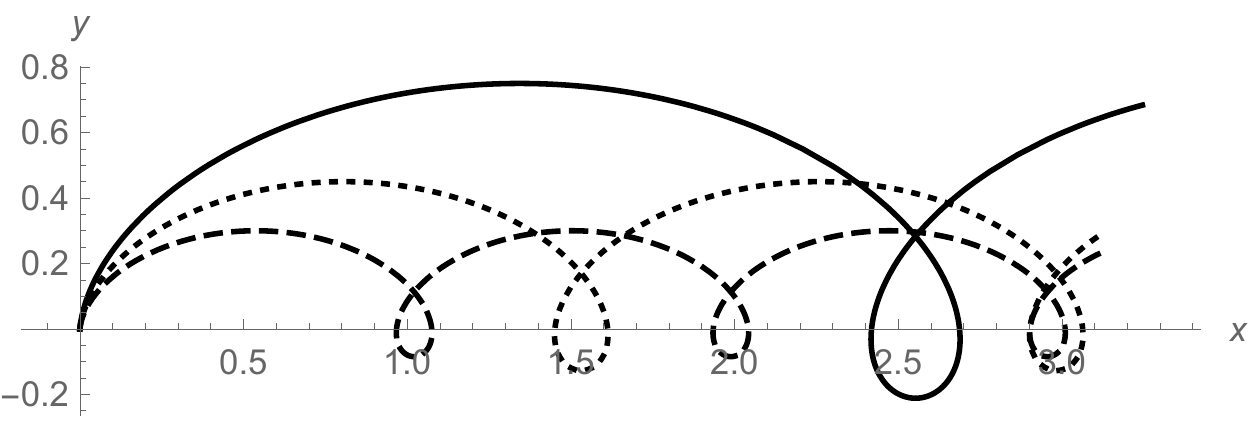}\\
({\bf a})&({\bf b})\\  
\end{tabular}
\caption{\small(\textbf{a}) particle trajectories in the 
$z=0$ plane for $0\le t\le
5$. Charge $q=1$, constant electric field $E=1$ in the positive $y$
direction, constant magnetic field $B=0.4$ 
in the positive $z$ direction.
Initial velocity ${\bf  v}_0$ = $(1,0,0)$.
Dashed line: ${\EE}=0.1$; dotted line: ${\EE}=0.3$; 
solid line: ${\EE}=0.7$; (\textbf{b}) Particle trajectories in the $z=0$ plane for $0\le t\le
5$. Charge $q=1$, constant electric field $E=1$ in the positive $y$
direction, constant magnetic field $B=1.6$ in the 
positive $z$ direction.
Initial velocity ${\bf  v}_0$ = $(0.1,0.995,0)$.
Dashed line: $\EE=0.2$; dotted line: $\EE=0.3$; solid line: 
${\EE}=0.5$.}
\label{fig2}
\end{figure}

Analytic solutions are easy to find for pure electric field or pure magnetic field. The solution for $B=0,$ satisfying the boundary conditions \equ{initial-cond-massless}, reads
\begin{equation}\label{sol_B=0}
\begin{array}{l}
x(t) = \dfrac{ v_{0x}}{\om}  
 \log\lp \dfrac{\sqrt{(\om t)^2+2v_{0y}\om t+1}+ \om t+ v_{0y}}{(1+v_{0y})}\rp  ,\es
y(t) = \dfrac{1}{\om}\lp \sqrt{(\om t)^2+2v_{0y}\om t+1}-1 \rp 
  , \es
z(t) = \dfrac{v_{0z}}{v_{0x}} x(t)   ,
\end{array}
\end{equation}
with $\om=qE/\EE$,
whereas the solution for $E=0$ with the same boundary conditions reads
\begin{equation}\label{sol_E=0}
\begin{array}{l}
x(t) = \dfrac{1}{\OM} 
\Lp  -v_{0y}(\cos(\OM t)-1)+ v_{0x}\sin(\OM t) \Rp   ,\es
y(t) = \dfrac{1}{\OM} 
\Lp v_{0x}(\cos(\OM t)-1)+ v_{0y}\sin(\OM t) \Rp   ,\es
z(t) = v_{0z} t   ,
\end{array}
\end{equation}
where $\OM=qB/\EE$.
We didn't find analytic solutions of the system 
\equ{eq-motion-constB,E-massive'} in the presence of both the electric and the magnetic fields, but a numerical analysis is
summarized in Figures \ref{fig1} and \ref{fig2},
where we have confined the movement to the plane $(x,y)$
by setting to zero
the initial velocity component $v_{0z}$. Figure \ref{fig1}
displays
the particle trajectory for four values of the ratio $B/E$: 
as expected, the one for $B>E$ is bounded in the $y$-direction, which is the direction of the electric field, and
exhibits a drift in the orthogonal direction. On the other hand, the two trajectories  for $B>E$ are unbounded in both directions.
The dotted line corresponds to the limiting case $B=E$, 
{which shows a similar behaviour.}
{In the case of}
$B=0$, we have the trajectory equation
\[
y(x)=-\frac{1}{\omega}+\frac{v_{0x}}{\omega}\cosh\lp\frac{x}{v_{0x}/\omega}+\mbox{sech}^{-1}(v_{0x})\rp,
\]
which is not  the catenary curved observed in the 
case of a massive particle~\cite{Landau,Bittencourt},
except if   $v_{0x}=1$. {See, in particular, p. 55   
of~}\cite{Landau}.
Figure \ref{fig2} displays the trajectories for three values of the 
total energy $\EE$, showing clearly the increase of the inertia
with increasing energy, for cases (a) of $B<E$ and (b) of $B>E$.

\unskip

\section{The Spinning Charged and Massless Particle} \label{spinning}
 We turn now to the case of a spinning 
 particle~\cite{balachandran,brink1,brink2,Fainberg1,Aliev1},
  completing action (\ref{action-spinless}) by terms 
  involving the spin degrees of freedom. The latter are 
  described by Grassmann odd (i.e.,~anti-commuting) variables: 
  a Lorentz vector $\psi^\mu(\la)$ and a scalar $\chi(\la)$. 
We restrict here to the less well established case of 
a massless particle. Recent accounts for the massive spinning particle may be found in~\cite{Rohrlich,Kosyakov-book}

  The manifestly
 Lorentz invariant action 
 reads, as an integral along a curve $\mathcal{C}$ 
 par\-am\-et\-riz\-ed by $\la$ {{(We follow the conventions of}~\cite{balachandran})}:  
\begin{equation}
	\begin{array}{l}
S = -\dint_{\!\!\!\mathcal{C}}d\lambda 
\Lp \frac{1}{2e}\dot{x}^{\mu}\left(\dot{x}_{\mu}-i\chi\psi_{\mu}\right)
+ \dfrac{i}{2}\psi^{\mu}\dot{\psi}_{\mu}
+  q A_{\mu}\dot{x}^{\mu}- \dfrac{iq}{2}e\psi^{\mu}F_{\mu\nu}\psi^{\nu}\Rp ,
	\end{array}
	\label{action4}
\end{equation}
where a dot means a derivative with respect to $\la$.

The action (\ref{action4}) is invariant, up to boundary terms,
 under arbitrary re\-par\-amet\-riz\-ations 
of $\lambda$ and local supersymmetric transformations.  
With $ \varepsilon(\lambda)$ (even) 
and $\alpha(\lambda)$ (odd) as infinitesimal parameters, 
these transformations read, respectively {({The second 
term in the transformation of }$\p_5$, which does not appear 
in~\cite{balachandran}, is necessary and may be found in {Equation} (6.2) 
of~\cite{brink2})},  
\begin{equation}
\begin{array}{l}
	\delta_{\varepsilon} x^{\mu}
	=\varepsilon \dot{x}^{\mu},\hspace{70pt} \\[3 mm]
	\delta_{\varepsilon} e = \dot{\varepsilon}e
	+\varepsilon\dot{e} ,\\[3 mm]
	\delta_{\varepsilon} \psi^{\mu}
	= \varepsilon \dot{\psi}^{\mu},\\[3 mm]
	\delta_{\varepsilon} \chi = \dot{\varepsilon}\chi 
	+  \varepsilon \dot{\chi},
		\end{array}
\begin{array}{l}
	\delta_{\alpha} x^{\mu}=i\alpha \psi^{\mu},\\[3mm]
	\delta_{\alpha} e =-i \alpha \chi ,\\[3mm]
	\delta_{\alpha} \psi^{\mu}
	=-\alpha\left(\dot{x}^{\mu}
	-\frac{i}{2} \chi\psi^{\mu}\right)/e,\\[3mm]
	\delta_{\alpha} \chi=2\dot{\alpha}.
	\end{array}
		\label{gaugeT}
\end{equation}

The electromagnetic potentials and fields then transform as 
\begin{equation}
\begin{array}{l}
	 \delta_{\varepsilon}A_{\mu}
	 =\varepsilon \dot{A}_{\mu} ,\\[3 mm]
	  \delta_{\varepsilon}F_{\mu\nu}=\varepsilon \dot{F}_{\mu\nu},
\end{array}
\begin{array}{l}
 \delta_{\alpha}A_{\mu}=i\alpha \partial_{\nu}A_{\mu}\psi^{\nu},\\[3 mm]
 \delta_{\alpha}F_{\mu\nu}=i\alpha \partial_{\rho}F_{\mu\nu}\psi^{\rho}.
	\end{array}
		\label{gaugeT-fields}
\end{equation}

The commutator of two supersymmetry transformation yields the combination of a reparametrization and of a supersymmetry
transformation:
	\begin{equation}
\lbrack \delta_{\alpha},\delta_{\beta}\rbrack  
=  \delta_{ \tilde{\varepsilon}}+ \delta_{\tilde{\alpha}},	\end{equation}
with their infinitesimal parameters
\begin{equation*}
 \tilde{\varepsilon}=\frac{2i}{e}\alpha\beta,\qquad
\tilde{\alpha}=-\frac{i\chi}{e}\alpha\beta.
\end{equation*}

  Varying the action with respect to $e(\la)$ and $\chi(\la)$ 
yields the two constraints 
\begin{equation}\label{constraint-bose}
C_1(\la) := \dfrac{\dot{x}^{\mu}\dot{x}_\mu}{e^2}
-i\dfrac{\chi\dot{x}^{\mu}\psi_{\mu}}{e^2}
+i q \psi^{\mu}F_{\mu\nu}\psi^{\nu} = 0,
\end{equation}
\begin{equation}\label{constraint-fermi}
C_2(\la) := \dot{x}^{\mu}\psi_\mu=0.
\end{equation}

{We observe that the $\chi$-term in the 
{bosonic constraint} \equ{constraint-bose}
vanishes due to {the fermionic constraint}~\equ{constraint-fermi}.}

The dynamical equations are obtained by varying the action with respect to $x^\m(\la)$ and $\p^\m(\la)$:
\begin{equation}\label{eqs-of-motion-spin}
\begin{array}{l}
\dfrac{d}{d\lambda}\left(\dfrac{\dot{x}_{\mu}}{e}
-i\dfrac{\chi\psi_{\mu}}{2e}\right)
-q \left(\dot{x}^{\nu}F_{\mu\nu}
-\dfrac{ie}{2}\psi^{\rho}
\partial_{\mu}F_{\rho\sigma}\psi^{\sigma}\right) = 0,\es
\dot{\psi}_{\mu}+\dfrac{\dot{x}_{\mu}\chi}{2e}
-e q F_{\mu\nu}\psi^{\nu} = 0.
\end{array}
\end{equation}

The local supersymmetry transformation \equ{gaugeT} for $\chi$ shows that the latter is a pure gauge of freedom, which will be set from now on to zero:
\begin{equation}\label{susy-gauge-fix}
\chi=0.
\end{equation}

This fixes the supersymmetry invariance. Later on, we will also fix 
reparametrization invariance by choosing a specific parametrization, 
namely $\la$ = $t$, instead of attributing a value to the
{\textit{einbein}} $e$ as is often 
done in  the literature~\cite{casalbuoni1}--\cite{geyer}.

Using the dynamical Equation \equ{eqs-of-motion-spin}, the anticomutativity of the 
$\p^\m$'s and {Equation} \equ{F=dA}, one shows that 
the left-hand sides of the 
constraints obey the equations
\begin{equation}\label{der-of-constr}
\dot C_1 = 0,\quad\quad \dfrac{\dot C_2}{C_2} = \dfrac{\dot e}{e}.
\end{equation}

These are consistency conditions that show that 
constraints \equ{constraint-bose}
and \equ{constraint-fermi}
are automatic consequences of the equations of motion 
\equ{eqs-of-motion-spin} if they are satisfied for 
some initial value $\la_0$ of the evolution parameter.

The theory  with Grassmann parameters just described is the
appropriate one for an Hamiltonian formulation and a subsequent 
quantization, as it has been done for the free particle 
in~\cite{brink2,vanholten} and in~\mbox{\cite{galvao-teitelboim,geyer}}  
with  electromagnetic interaction, but only in the massive
case.
A theory easier to interpret as the one of a classical 
spinning particle may be obtained introducing  the spin tensor $\S$, whose~components are even Grassmann 
numbers~\cite{balachandran,Kosyakov-book}:
\begin{equation}\label{def-spin}
\S_{\m\n} = -i\p_\m\p_\n = -\S_{\n\m}.
\end{equation}

This formulation is the one that is suitable as an effective
theory, which ought to describe the semi-classical limit
of the quantum theory in terms of expectation values.

The  constraints \equ{constraint-bose} and \equ{constraint-fermi}
now read
\begin{eqnarray}
&&\dfrac{\dot{x}_{\mu}\dot{x}^\mu}{e^2}  
- q F^{\mu\nu}\S_{\m\n} =0,
\label{cov-cons-bose}\es
&&\dot{x}^{\mu}\S_{\mu\nu} = 0,
\label{spin-constraint}\end{eqnarray}
and  the equations of motion \equ{eqs-of-motion-spin} 
take now the form
\begin{eqnarray}
&&\dfrac{d}{d\lambda}\left(\dfrac{\dot{x}_{\mu}}{e}\right)
-q \left(F_{\mu\nu}\dot{x}^{\nu}
+\dfrac{e}{2}\partial_{\mu}F^{\rho\sigma}\S_{\rho\s}\right)= 0,
\label{pos-eq}\es
&&\dot{\S}_{\m\n} 
- q\,e(F_\m{}^\s\S_{\s\n} - F_\n{}^\s\S_{\s\m} ) = 0.
\label{spin-eq}\end{eqnarray}

\subsection{Time Parametrization}

Choosing  now  the time parametrization, $\la$ = $t$, we see that
the spin constraint \equ{constraint-fermi} can be solved for
 the component $\p_0$ in terms of the $\p_i$ ($i=1,2,3$):
\begin{equation}\label{psi0}
 \p_0 =  -\dot{x}^i\p_i,
\end{equation}
where a dot now means the time derivative.

Instead of working with the odd Grassmann variables $\p_\m$, we shall
use the even Grassmann spin tensor $\S$ defined in 
{Equation} \equ{def-spin}. 
Its components can be written as components of the two 3-vectors
\begin{equation}\label{vectors-s-n}
{\bf n}=(\S_{01},\S_{02},\S_{03}) ,\quad 
{\bf s}=(\S_{23},\S_{31},\S_{12}) ,
\end{equation}
so that the spin constraint \equ{spin-constraint} can be solved for 
$\mathbf{n}$ in term of  $\mathbf{s}$:
\vspace{-6pt}

\begin{equation}\label{n=n(s)}
\mathbf{n} = \dot{\mathbf{x}}\times\mathbf{s},
\end{equation}
and we observe that the vector $\mathbf{n}$ 
is orthogonal to the velocity. We remark that 
{Equation} \equ{spin-constraint} (or \equ{n=n(s)}) is identical to 
the covariant version of the Frenkel 
condition~\cite{Frenkel1,Frenkel2,Kosyakov-book}---introduced for 
the massive case!---that the 3-vector $\bf n$ vanishes 
in the rest frame of the particle. 

The constraint \equ{cov-cons-bose} and the dynamical 
equations for the position  \equ{pos-eq} 
read, respectively,
\begin{equation}\label{constraint-bose-class-spin}
\dfrac{1-\dot{\mathbf{x}}^2}{e^2} 
+ 2q(\mathbf{s}\cdot\mathbf{B} + \mathbf{n}\cdot\mathbf{E}) = 0,
\end{equation}
and
\vspace{-6pt}

\begin{equation}\label{eq-motion-with spin}
\begin{array}{l}
\dfrac{d}{dt}\lp\dfrac{1}{e}\rp- q \mathbf{E} \cdot\dot{\mathbf{x}} 
+q\,e(\mathbf{s}\cdot\pa_t\mathbf{B}+\mathbf{n}\cdot\pa_t\mathbf{E})
= 0,\\[4mm]
\dfrac{d}{dt}\lp\dfrac{\dot{\mathbf{x}}}{e}\rp 
- q(\mathbf{E} +  \dot{\mathbf{x}} \times \mathbf{B}) 
-q\, e\sum_{i=1}^3 (s_i\nabla B_i+n_i\nabla E_i)
= 0.
\end{array}
\end{equation}
In the same way, {the Equation}  \equ{spin-eq} for the spin 
vector $\mathbf{s}$ reads
\begin{equation}\label{3D-spin-eq}
\dot{\mathbf{s}} + q\, e 
(\mathbf{E}\times\mathbf{n} + \mathbf{B}\times\mathbf{s}) = 0,
\end{equation}
 and the one for $\mathbf{n}$, $\dot{\mathbf{n}} + q\, e 
(\mathbf{B}\times\mathbf{n} -\mathbf{E}\times\mathbf{s}) = 0$,
 following from {Equations}
\equ{n=n(s)} and \equ{eq-motion-with spin}.

{{\bf N.B.} We observe that the {\textit{einbein}} variable $e(t)$
is dynamical, its evolution being defined by the first of 
{Equation} \equ{eq-motion-with spin}. On the other hand, the constraint
\equ{constraint-bose-class-spin} fixes the absolute value of
the velocity, which~may thus be variable and
 different from that of the light.}
This feature is a peculiarity of the massless theory.
 We will check this for some concrete examples in Section 
 \ref{const-field-spin}.

\subsection{{Conservation Laws}}

{The first obvious conservation law is that of electric charge: the charge $q$ of the particle is just a parameter of the action
defining the magnitude of the coupling with the external electromagnetic field.}

{Energy, momentum and angular momentum conservations, although not as trivial, are easily derived applying the Noether procedure. Their validities depend on the symmetries preserved by the external electromagnetic field.}

{In the case of a stationary exterior field, with
$\pa_t A_\m=0$, integration of 
the first of {Equation} \equ{eq-motion-with spin} leads to
the same conserved energy $\EE$ as in the spinless case:
\begin{equation}\label{energy-spin}
\EE=\dfrac{1}{e(t)} + q A_0(\mathbf{x(t)}).
\end{equation}}

Invariance under the space translations holds if the electromagnetic fields $\bf{E}$ and $\bf{B}$ are also constant in space, which is assured by the 4-potential vector
$A_\m$ = $f_{\m\n}x^\n$, where the $f_{\m\n}$s are constants,
antisymmetric in $\m,\,\n$.~Then, this implies that the conservation of all four components of the energy-momentum 4 vector:
\begin{equation}\label{momentum}
\PP_\m = \dfrac{1}{e(t)}\dot{x}_\m(t)-qf_{\m\n}x^\n(t).
\end{equation}

Finally, we have rotation invariance in a plane if the electric field vanishes and the magnetic field $\bf B$ points to a fixed direction. Choosing this direction to be the $z$ -axis, we deduce the conservation of the $z$ component of the total angular momentum:
\begin{equation}\label{ang-momentum}
J_z = -\dfrac{1}{e(t)}(\dot{x}(t) y(t) - \dot{y}(t) x(t)) + \dfrac{qB}{2}(x^2(t)+y^2(t)) + s_z(t),
\end{equation}
the last term representing the spin part $s_z=- i\p^1\p^2$ (see definitions \equ{def-spin} and 
\equ{vectors-s-n}).

\subsection{Physical Interpretation of the Classical Theory}

In order to be able to interpret the theory as a truly classical
one, in terms of real numbers, one should forget about the Grassmann character of the spin variables $\S_{\m\n}$ (or $\mathbf{s}$
and $\mathbf{n}$) and consider them as real number quantities. 
The theory would still be defined by the set of {Equations} 
(\ref{cov-cons-bose})--(\ref{spin-eq}),
or~(\ref{n=n(s)})--(\ref{3D-spin-eq}) in the 3D notation. These equations do no more
derive from an action principle, so that their consistency must be checked. Unfortunately, it happens that the spin constraint
\equ{spin-constraint} is incompatible with the rest of the equations. Indeed, deriving it with respect to the evolution parameter $\la$, \begin{equation}\label{evol-spin-constr}
\dfrac{d}{d\la}\lp \dot{x}^{\mu}\S_{\mu\nu}\rp =
\dfrac{q}{2} \pa^\m F^{\rho\s} \S_{\m\n}\S_{\rho\s},
\end{equation}
which only vanishes for special field configurations, such as, 
e.g.,  a constant 
electromagnetic field ({{If~the} $\S^{\m\n}$s still 
were even Grassmann numbers, products of odd elements as 
in {the definition} \equ{def-spin},
then the expression $\S^{\m\n}\S^{\rho\s}$ would be 
antisymmetric in the three indices $\m,\rho,\s$ and the 
right-hand side of {Equation}
\equ{evol-spin-constr} would {vanish due to} $F_{\rho\s}$
= $\pa_\rho A_\s-\pa_\s A_\rho$)}, 
which we shall consider
in Section \ref{const-field-spin}. 

An alternative could be to use the constraint and equations of motion
\equ{cov-cons-bose}, the spin-constraint \equ{spin-constraint} or
\equ{n=n(s)} on one hand, and the spin Equation \equ{spin-eq} 
only for $\m\n$ = $ij$, i.e., for the spin 3-vector $\bf{s}$,
on the other hand. 
However, such a choice would break Lorentz covariance.

\subsection{Constant Electromagnetic Field}\label{const-field-spin}

As we saw in the last subsection, the restriction to a constant electromagnetic field preserves the full set of
the Lorentz covariant constraints and dynamical equations.

We shall consider the same configuration with a constant 
electromagnetic field as
discussed in the spinless case at the end of Section \ref{example_spinless}, i.e., with
$\bf{E}$ = $(0,E,0)$ and $\bf{B}$ = $(0,0,B)$, 
with mass zero 
and with the time parametrization, $\la=t$.  
The equations of motion for $e$, $x$, $y$ and $z$ take the same 
form \equ{eq-motion-massless}, or {their} integrated form
\equ{eq-motion-constB,E-massive'}, as in the spinless case. The
conserved energy is given by {Equation} \equ{energy'}. The~constraint
equation and the spin equations read 
 \begin{equation}\label{constr-constant-fields}
\dfrac{1-\dot{\mathbf{x}}^2}{e^2}
+2q\left(B s_z+ E\dot{z} s_x-E\dot{x} s_z\right))=0,
\end{equation}
\begin{equation}\label{eq-constant-fields}
\begin{array}{l}
\dot{s}_x
=q\, e \left(E \dot{y} s_x- E \dot{x} s_y +B s_y\right),\\[3 mm]
\dot{s}_y= -q\, e  B s_x,\\[3 mm]
\dot{s}_z=q\, e \left(E \dot{y} s_z- E \dot{z} s_y \right).
\end{array}
\end{equation}

Solutions for $B=0$ or $E=0$ are easy to find. 
We use again the notations
\begin{equation}
\om = \dfrac{qE}{\EE},\quad\quad \OM = \dfrac{qB}{\EE},
\eqn{omega-Omega}
of Section \ref{example_spinless}
and the boundary conditions \equ{initial-cond-massless}.

For $B=0$, the evolution of the position coordinates is given 
by {Equation} \equ{sol_B=0}, but with differences in the initial velocity
components due to the modified constraint 
(see {Equation} \equ{constraint-B=0}).
The evolution of the spin components is given by
\begin{equation}\label{sol-spin-B=0}
\begin{array}{l}
s_x(t) = -\dfrac{s_{0y}v_{0x}(v_{0y}+\om t )}{1-v_{0y}^2} 
+s_{0x}\sqrt{1+2v_{0y}\om t+\om^2t^2},\es
s_y(t) = s_{0y},\es
s_z(t) = -\dfrac{s_{0y}v_{0z}(v_{0y}+\om t )}{1-v_{0y}^2} 
+s_{0z}\sqrt{1+2v_{0y}\om t+\om^2t^2}.\es
\end{array}
\end{equation}

The constraint reads
\begin{equation}\label{constraint-B=0}
 \EE(1-v_{0x}^2-v_{0y}^2-v_{0z}^2)
 +2\omega (s_{0x}v_{0z}-s_{0z}v_{0x}) = 0.
\end{equation}

In view of its constancy (see the first of {Equation} 
\equ{der-of-constr}), we have taken it at $t=0$: it is thus 
a constraint on the initial parameters $v_{0i}$ and $s_{0i}$.
One sees that the particle's velocity is not constrained to be 
equal to  the velocity of light $c=1$---excepted for very peculiar
initial spin/velocity configurations in which the spin part of 
{Equation} \equ{constraint-B=0} vanishes. The particle's velocity can even 
exceed $c$. We note that Lorentz invariance remains nevertheless 
unbroken. This feature, peculiar to the present approach of
the classical massless spinning particle, will be encountered in
various other examples, as we will~see.

For the purely magnetic case, $E=0$,  the position coordinates are given 
by {Equation} \equ{sol_E=0}, and the spin components by the precession equations
\begin{equation}\label{sol-spin-E=0}
\begin{array}{l}
s_x(t) = s_{0x} \cos(\Omega t)+s_{0y} \sin(\Omega t),\es
s_y(t) = s_{0y} \cos(\Omega t)-s_{0x} \sin(\Omega t),\es
s_z(t) = s_{0z},
\end{array}
\end{equation}
where $(s_{0x},\,s_{0y},\,s_{0z})$ is the spin vector at $t=0$.
The constraint reads
\begin{equation}\label{constraint-E=0}
 \EE(1-\dot{\mathbf{x}}^2)+2\Omega  s_{0z} = 0.
\end{equation}

One sees that the magnitude of the particle's
velocity---which here is constant due to 
the first of {Equation} \equ{der-of-constr}---can be higher 
or lower than the
velocity of light, depending on the sign of $\OM s_{0z}/\EE$.

In the case of both $E$ and $B$ being non-zero, one has first to
solve the constraint \equ{constr-constant-fields} for one of the velocity components, let us say the $x$-component. In view of the constancy
of the constraint (see the first of {Equation} \equ{der-of-constr}), 
it is sufficient to do it at the initial time $t=0$ for the 
initial velocity component $v_{0x}$. It~is thus a quadratic 
equation for the initial
value of the velocity component, $v_{0x}$.
In order to obtain real solutions, the discriminant
\begin{equation}\label{discriminant}
\D:=(1-v_{0y}^2-v_{0z}^2)\EE^2 
+ 2(s_{0x}v_{0z}\om + s_{0z}\OM)\EE + s_{0z}^2\om^2
\end{equation}
must be non-negative, hence the reality condition,
\begin{equation}\label{reality-con}
\D\ge 0,
\end{equation}
must hold. 

In order to be more explicit, we specialise from now on to the case 
of trajectories in the $(x,y)$-plane with the spin pointing to the 
$z$-direction, which is guaranteed by the initial conditions
\begin{equation}\label{plane-bound-cond}
v_{0z}=0,\quad\quad s_{0x}=s_{0y}=0.
\end{equation}

The reality condition holds if and only if
\begin{equation}\label{ineq-v0y}
v_{0y}^2\le 1 +\dfrac{2\OM}{\EE} s_{0z} +
\dfrac{\om^2}{\EE^2} s_{0z}^2.
\end{equation}

A necessary condition for this inequality is the positivity of 
the right-hand side, which holds in the following three cases:
\begin{equation}\label{planar-reality-cond}
\begin{array}{l}
\mbox{(a)  }|\om| \ge |\OM| ,\quad\quad\quad \forall\, s_{0z},\es
\mbox{(b)  }|\om| < |\OM| ,\quad\quad \quad
s_{0z} < -\dfrac{\EE\OM}{\om^2}
      -\dfrac{\EE^2}{\om^2}\sqrt{\dfrac{\OM^2-\om^2}{\EE^2}},\es
\mbox{(c)  }|\om| < |\OM| ,\quad\quad \quad
s_{0z} > -\dfrac{\EE\OM}{\om^2}
      +\dfrac{\EE^2}{\om^2}\sqrt{\dfrac{\OM^2-\om^2}{\EE^2}}.\es
\end{array}
\end{equation}

Figure \ref{fig4} shows some characteristic solutions.
\begin{figure}[htb]
\begin{subfigure}{.5\textwidth}
  \centering
  \includegraphics[scale=0.5]{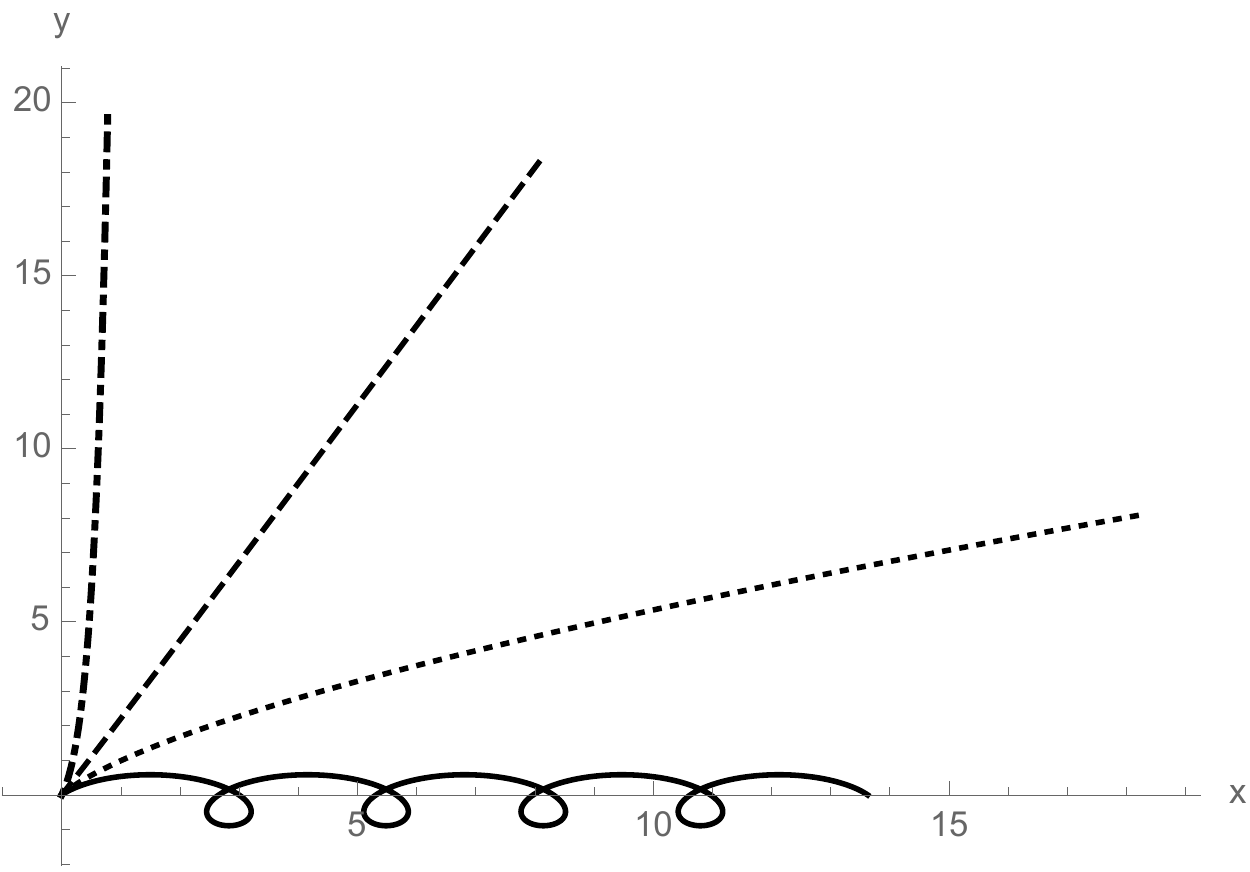}
  \caption{Trajectories in the $z=0$ plane.}
  \label{fig4a}
\end{subfigure}
\begin{subfigure}{.5\textwidth}
  \centering
  \includegraphics[scale=0.5]{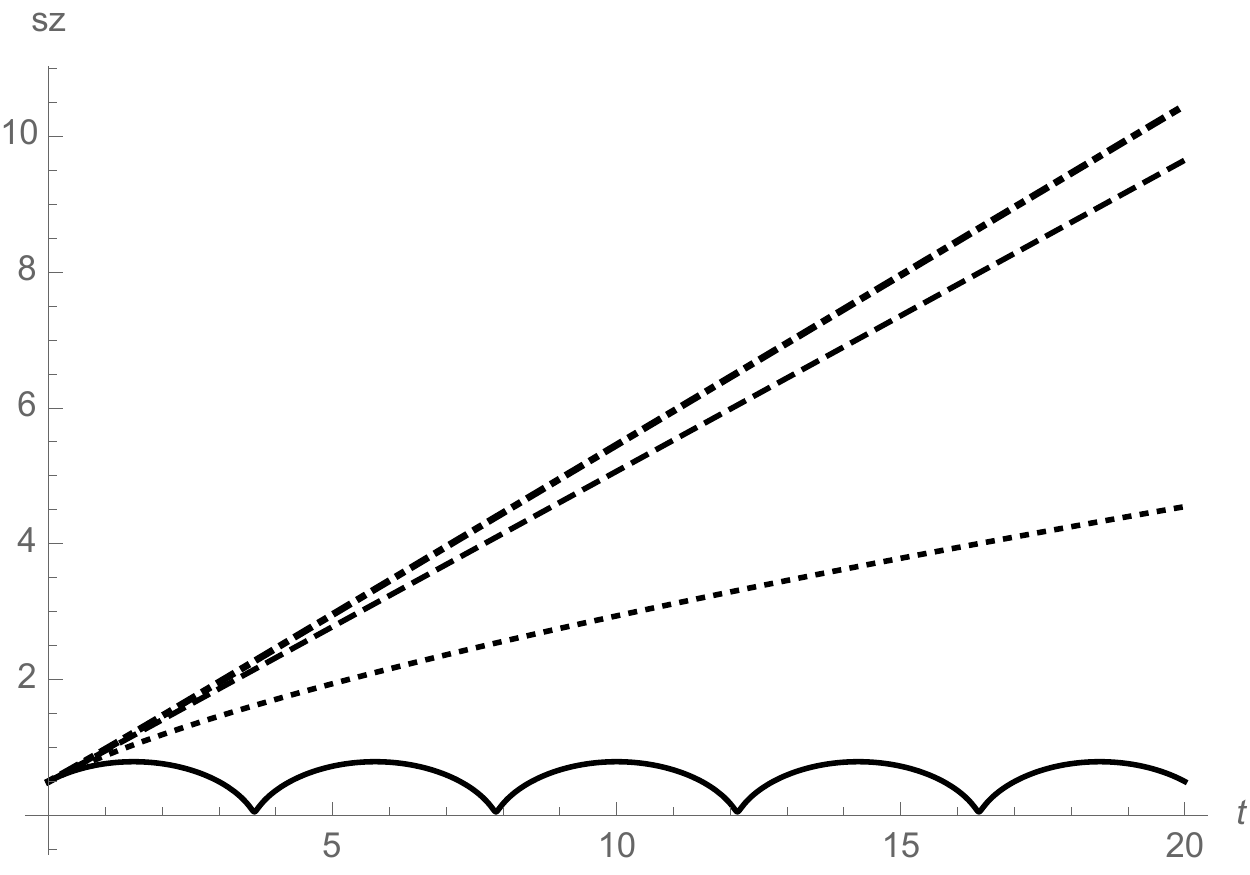}
  \caption{Spin $s_z(t)$.}
  \label{fig4b}
\end{subfigure}
\begin{subfigure}{.5\textwidth}
  \centering
  \includegraphics[scale=0.5]{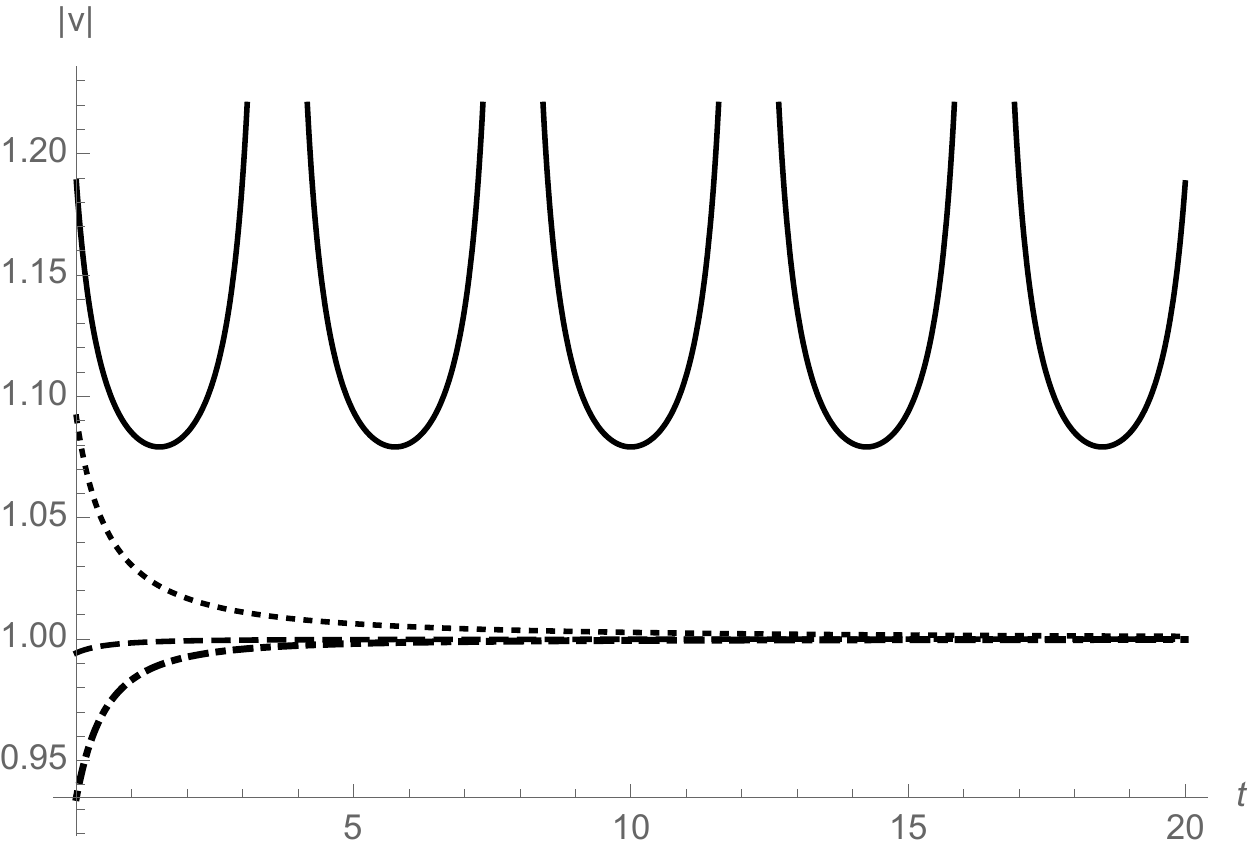}
  \caption{Velocity $|\bf v(t)|$.}
  \label{fig4c}
\end{subfigure}
\caption{\small {(a)} Particle trajectories in the $z=0$ plane, 
{(b)} spin $s_z(t)$  
and {(c)} velocity $|\bf v|$ = $|\dot{\mathbf{x}}(t)|$ for a constant 
electric field $E$ in the positive $y$
direction and a constant magnetic field $B$ in the positive $z$ direction.
 Parameters' values are chosen as:
charge $q=1$, energy $\EE=2$, initial spin ${\bf s}(0)$ =
$(0,0,0.5)$ and initial velocity 
${\bf v}(0)$ = $(v_{0x},0.9,0)$, $v_{0x}$ being the largest of the solutions of the constraint \equ{constr-constant-fields}. The~following field configurations have been chosen:
 $B=3.2,\, E=2$ (solid lines); $B=E=2$ (dotted lines);
 $B=0.8,\, E=2$ (dashed lines); $B=0,\, E=2$ (dotted-dashed lines).
The~$B=0$ trajectory would be on the upper vertical
axis in the case of ${\bf v}(0)$ = $(v_{0x},1,0)$.
}\label{fig4}\end{figure}
One observes the same behaviour as in the spinless 
case for the trajectories
(Figure \ref{fig4}a): bounded in the electric field direction (component
$\dot{y}(t)$ for 
$|B|>|E|$, unbounded for $|B|\le|E|$. A similar behaviour happens 
for the spin, as shown in Figure \ref{fig4}b: $s_z(t)$ is unbounded for 
$|B|\le|E|$. In fact, all numerical examples investigated show this transition between bounded and unbounded behaviour 
happening both for $\dot{y}$ and $s_z$ at $|B|=|E|$. In Figure \ref{fig4}c,
one sees the variation of the velocity's absolute value in function of $t$. This velocity turns out to be always bounded.

\section{Discussion and Conclusions}

{Our treatment of the massless charged particle
in interactions with an external electromagnetic field led to two results, which, to the best of our knowledge, are new. 
One of them is the proper definition of energy 
given in {Equation} \equ{en-cons-massless} 
for the spinless particle 
and in {Equation} 
\equ{energy-spin} for the spinning one. In the time pametrization, 
the inverse of the \textit{einbein}
function, $1/e(t)$, plays the role of the ``kinetic energy''.
This~was not recognized in the previous literature, where
parametrization invariance is usually invoked in order 
to fix to some value 
the arbitrary function $e(\la)$.
}

{The second result has been a non-constant particle's 
absolute velocity,
$|\mathbf{v}|$, different from $c$ in particular. It is interesting 
to compare with the massive particle case. For the massless particle,
the~result comes from the spin-electromagnetic field interaction present
in the {constraint}  \equ{constraint-bose}. 
In~the massive case, choosing the proper time 
parametrization ({{Dots now mean proper time 
derivative} $d/d\tau$}),   
where $\dot{x}^2=1$, we can solve this constraint for $e(\la)$:
\[
\dfrac{1}{e}=m\sqrt{1+q\S^{\m\n}F_{\m\n}/m^2},
\]
with $\S^{\m\n}$ defined by {Equation} \equ{def-spin}. 
Combining with the energy equation
\[
\EE = \dfrac{\dot{t}}{e}+qA_0,
\]
(deduced from the first of {Equation} \equ{eqs-of-motion-spin} for
$\m=0$ in the stationary case), we obtain
the time delay~equation
\begin{equation}\label{time-delay}
\dfrac{d\tau}{dt} = 
\dfrac{m}{\EE-qA_0}\sqrt{1+q\S^{\m\n}F_{\m\n}/m^2},
\end{equation}
a result similar to the one of~\cite{vanholten-el-dyn},
but not quite equal due to a different set of dynamical equations.
An~alternative theory for the massive particle has been very recently
proposed in Refs.~\cite{deriglazov1,deriglazov2}, where a speed limit different from $c$ has been found, a result reinterpreted as a modification of the space-time metric. 
As said, these latter results are valid for the massive theory. The essential difference between the massive and the non-massive theories lies in the fact that, in the former case, the constraint may be solved for the 
{\textit{einbein}} 
variable $e$, whereas, in the latter case, $e$ remains a dynamical variable---interpreted as the kinetic energy---and the constraint gives a (non-trivial) relation between the velocity components, as we have illustrated in specific examples. 
}

We have considered both the pseudo-classical supersymmetric
theory with odd Grassmann parameters, suitable for a 
canonical quantization, and the classical theory with spin 
described by real valued functions, which we have argued 
to better describe the classical limit of the quantum theory 
in terms of expectation values. The drawback
of the latter description is the absence of an action 
principle and the
incompatibility of the full system of equations excepted 
for special external field configurations, such as a 
constant one. 

It is for a constant field configuration that 
we have calculated explicit 
solutions showing the characteristic behaviours of the particle, 
in particular the fact that, due to the interaction of the 
spin with the external field, its velocity is in general 
different from the velocity of light, without contradiction 
{to} Lorentz invariance. 

{Finally, let us comment that a speed higher than $c$ would of course generate a conflict with causality, as
tachyons do, and may constitute an argument explaining 
the absence of such particles in the realm of fundamental 
physics. On the other hand, there would be no such problem 
in applications to condensed matter physics, such as graphene. There, 
the critical velocity $c$, taken equal to 1 in 
this paper and which plays the role of the relativistic 
``velocity of light'', is far smaller~\cite{graphene1,graphene2} 
than the true velocity of light in the vacuum, 
the one which defines the causal structure of space-time.
}

{As announced in the Introduction, our considerations 
are purely classical. Of course, some 
``correspondence principle'' should hold
such that, at the level of the quantum theory, 
the behaviour of the expectation values, e.g.,  of 
the velocity, should, in certain limiting cases, 
follow the classical behaviour discussed here. 
However, a check of this conjecture 
lies beyond the scope of the present~work.
}



\subsubsection*{Acknowledgements}
We would like to thank Alexei Deriglazov for 
pointing out to us the recent and interesting works of 
Refs.~\cite{deriglazov1,deriglazov2}.
This work was partially funded by the
 Conselho Nacional de Desenvolvimento Cient\'{\i}fico e
 Tecnol\'{o}gico---CNPq, Brazil (I.M., Z.O.  and O.P.),
to the Coordena\cao\ de Aperfei\c coamento de Pessoal de N\ii vel 
Superior---CAPES, Brazil (I.M. and B.N.)
and to the Grupo de Sistemas Complejos de la Carrera de Física de la {Universidad Mayor de San Andr\'es, UMSA}, for their 
support (Z.O.).
B. N. would like to show his gratitude to Camila G. S. Moraes who immensely supported him during the course of this research.



\end{document}